\documentclass[%
 reprint,
 amsmath,amssymb,
 aps,
floatfix,
]{revtex4-2}

\usepackage[utf8]{inputenc}
\usepackage[T1]{fontenc}
\usepackage{graphicx}
\usepackage{multirow}%
\usepackage{array}
\usepackage{bm}
\usepackage{amsmath,amssymb,bbm}
\usepackage[normalem]{ulem}
\usepackage{hyperref,float}
\usepackage{xcolor}
\usepackage{theorem}
\usepackage{multirow}
\usepackage{cases}

\newcommand{\sfa}{\mathsf{a}}
\newcommand{\sfb}{\mathsf{b}}

\newcommand{\sfab}{\mathsf{ab}}
\newcommand{\sfo}{\mathsf{0}}
\newcommand{\ma}{m_\mathsf{a}}
\newcommand{\mb}{m_\mathsf{b}}
\newcommand{\mab}{m_\mathsf{ab}}
\newcommand{\mo}{m_\mathsf{0}}
\newcommand{\Ma}{M_\mathsf{a}}
\newcommand{\Mb}{M_\mathsf{b}}

\newcommand{\thetaa}{\theta_\mathsf{a}}
\newcommand{\thetab}{\theta_\mathsf{b}}

\newcommand{\ER}{Erd\H{o}s-R\'{e}nyi }

\newcommand{\add}[1]{{\color{black}#1}}

{\theorembodyfont{\upshape}
}

\newcommand{\vecm}{{\bf m}}

\newcommand{\vecrho}{{\boldsymbol \rho}}

\newcommand{\papertitle}{Diffusion dynamics of competing information on networks}

\begin{document}

\title{\papertitle}

\author{Teruyoshi Kobayashi}
\email{kobayashi@econ.kobe-u.ac.jp}

\affiliation{Department of Economics, Kobe University, Kobe, Japan \\
Center for Computational Social Science, Kobe University, Kobe, Japan}

\begin{abstract}%
Information diffusion on social networks has been described as a collective outcome of threshold behaviors in the framework of threshold models.
However, since the existing models do not take into account individuals' optimization problem, it remains an open question what dynamics emerge in the diffusion process when individuals face multiple (and possibly incompatible) information.
Here, we develop a microfounded general threshold model that enables us to analyze the collective dynamics of individual behavior in the propagation of multiple information.
The analysis reveals that the virality of competing information is fundamentally indeterminate. 
When individuals maximize coordination with neighbors, the diffusion process is described as a saddle path, thereby leading to an unpredictable symmetry breaking.
When individuals' choices are irreversible, there is a continuum of stable equilibria where a certain degree of social polarization takes place by chance. 
\end{abstract}

\maketitle

\section{Introduction}

New technologies, rumors, and political opinions occasionally spread globally through social ties among individuals. 
The dynamical processes of complex contagions have been extensively studied within the framework of threshold models to understand whether and to what extent a ``social meme'' (e.g., a particular technology, opinion, etc) spreads on a social network~\cite{Watts2002,Gleeson2007,gleeson2013binary,Nematzadeh2014,Brummitt2015PRE,kobayashi2015trend,bottcher2017critical,gleeson2018message,unicomb2019reentrant}. 

However, it is common in reality that multiple memes are competing each other, and the popularity of one meme often affects the virality of another;
examples include ``format wars'' (e.g., VHS vs Betamax, Blu-ray Disc vs HD-DVD, etc)~\cite{cusumano1992strategic,anscombe2008after}, political campaign (e.g., democrat/republican)~\cite{conover2012partisan,metaxas2012social,ferrara2017contagion,vasconcelos2019consensus,baumann2020modeling,cinelli2021echo}, and vaccination behavior (i.e., pro- and anti-vaccination)~\cite{johnson2020online,prieto2021vaccination,adepoju2021africa}. 
In some cases, only one meme survives (e.g., VHS and Blu-ray Disc), while in other cases, multiple memes coexist persistently.
The interplay between competing social memes  that takes place at both local and global scales thus plays a key role in understanding the actual diffusion dynamics.
In the context of \emph{simple contagion}, in which infection probability is given by a constant, spreading dynamics of two competing viruses/pathogens have been well studied~\cite{newman2005threshold,poletto2013host,van2014domination}.
In contrast, in the literature of complex social contagion, it is still unknown when and how individuals collectively spread multiple memes as a result of optimization behavior.

Here, we develop a generalized threshold model of global cascades that allows us to describe the propagation dynamics of competing memes.
Our model is ``microfounded'' in the sense that individual behavior is optimized; individuals maximize coordination with their neighbors.
In this model, therefore, any stationary state of the dynamical process, if it exists, is interpreted as a collective outcome of individuals' strategic choices, namely, a Nash equilibrium~\cite{morris2000contagion,jackson2007diffusion,jackson-zenou2015games,kobayashi2021dynamics,kobayashi2021unstable}.

Game theorists have long studied diffusion on networks that arises from strategic interactions between individuals connected by social ties~\cite{Jackson2008book,Easley2010book,jackson2011overview,jackson-zenou2015games,tabasso2019diffusion}. A pioneering work by Morris~\cite{morris2000contagion} studies a class of $2\times 2$ coordination games on regular graphs and derived a contagion threshold of the payoff parameter.
In the literature on network games~\cite{Jackson2008book,jackson-zenou2015games}, however, most of the studies focus on the equilibrium property rather than the dynamics of diffusion~\cite{jackson2007diffusion,ballester2006,chen2018AEJmultiple}.  
In studies of complex contagion in network science, on the other hand, individual behavior is often captured by a presumed threshold rule~\cite{Granovetter1978,Watts2002}. 
Unless individuals' optimization is taken into account, however, any extension of the threshold rule would be inevitably arbitrarily since there is no fundamental principle behind the rule.
In the current work, we provide a framework in which individual behavior is disciplined through coordination games.
Based on the game-theoretic approach, we endogenously obtain generalized threshold rules with which individuals decide whether to accept memes given the influence from others.

\section{A threshold model of cascades with competing memes}

Recently, it is shown that the (fractional) threshold rule used in the Watts cascade model~\cite{Watts2002} and the optimal strategy in a model of coordination games on networks~\cite{morris2000contagion,Jackson2008book,jackson-zenou2015games} are functionally equivalent~\cite{kobayashi2021dynamics}.
This indicates that a global cascade may be interpreted as a collective outcome of individuals' optimization behavior that maximizes their payoffs from coordination. 
However, while this equivalence provides a microfoundation for the Watts threshold model, the argument is limited to the case where individuals face a binary choice problem (e.g., cooperate or not cooperate, being active or inactive).
In this section, we aim to generalize the binary threshold rule by introducing a non-binary coordination games.

\subsection{Coordination game with two types of social memes}

We consider two types of social memes, respectively labeled as $\sfa$ and $\sfb$.
The memes can be either complementary, exclusive  or neutral. 
Each individual decides whether to accept $\sfa$ or $\sfb$, or both (called the \emph{bilingual option}, denoted by $\sfab$), referring to the popularity of each meme among local neighbors~\cite{oyama2015bilingual,kobayashi2021unstable}. 
Let $S \equiv \{\sfo,\sfa,\sfb,\sfab\}$ be the set of pure strategies where $s=\sfo$ indicates the status-quo (i.e., neither meme is accepted).  
In an infinitesimal time interval $dt$, randomly selected individuals update their strategies (i.e., \emph{asynchronous update}~\cite{gleeson2013binary,Melnik2013}) to maximize the payoffs of coordination games.
The payoff matrix for a bilateral coordination game is presented in Tab.~\ref{tab:two-good_payoff}.
\begin{table}[tb]
    \centering
    \caption{Payoff matrix of a coordination game. We assume $a,b>c>0$. The two memes are complementary (resp. exclusive) when $\tilde{c}<c$ (resp. $\tilde{c}>c$). }
    \begin{tabular}{cccccc}
     \multicolumn{1}{c}{}   &  &  \multicolumn{1}{c}{$\sfo$}    & \multicolumn{1}{c}{$\sfa$} & \multicolumn{1}{c}{$\sfb$}&\multicolumn{1}{c}{$\sfab$}  \\
    \cline{3-6}
    &  \multicolumn{1}{c}{$\sfo$} &\multicolumn{1}{|c}{$0,0$} & \multicolumn{1}{|c}{$0,-c$} & \multicolumn{1}{|c}{$0,-c$} &\multicolumn{1}{|c|}{$0,-2\tilde{c}$}\\
    \cline{3-6}
    &  \multicolumn{1}{c}{$\sfa$} &\multicolumn{1}{|c}{$-c,0$} & \multicolumn{1}{|c}{$a-c,a-c$} & \multicolumn{1}{|c}{$-c,-c$} &\multicolumn{1}{|c|}{$a-c,a-2\tilde{c}$}\\
    \cline{3-6}
      &  \multicolumn{1}{c}{$\sfb$} &\multicolumn{1}{|c}{$-c,0$} & \multicolumn{1}{|c}{$-c,-c$} & \multicolumn{1}{|c}{$b-c,b-c$} &\multicolumn{1}{|c|}{$b-c,b-2\tilde{c}$}\\
    \cline{3-6}
      &\multicolumn{1}{l}{\multirow{2}*{$\sfab$}} &\multicolumn{1}{|c}{\multirow{2}*{$\:\:\quad -2\tilde{c},0\:\: \quad$}} & \multicolumn{1}{|c}{\multirow{2}*{$a-2\tilde{c},a-c$}}&\multicolumn{1}{|c}{\multirow{2}*{$b-2\tilde{c},b-c$}}&\multicolumn{1}{|c|}{\multirow{2}{2.1cm}{$a+b-2\tilde{c}$,\\ ${}\qquad a+b-2\tilde{c}$}}\\
      & &\multicolumn{1}{|c}{} &\multicolumn{1}{|c}{} &\multicolumn{1}{|c}{} &\multicolumn{1}{|c|}{} \\
    \cline{3-6}
    \end{tabular}
    \label{tab:two-good_payoff}
\end{table}

Each element of the payoff matrix shows the returns for the corresponding strategy pair.
For instance, the pair $(-c,0)$ in the $(2,1)$th element of the matrix indicates that a player accepting meme $\sfa$ receives the payoff $-c$ while the other player receives $0$ by staying in the status quo.   
$a$ and $b$ are the benefits of coordinating with neighbors in adopting strategies~$\sfa$ and $\sfb$, respectively, and $c$ denotes the \add{fundamental} cost of accepting a meme, where we assume that $a,b>c>0$.
For example, two close friends having PCs will be better off using a common operating system rather than different ones.
$a,b>c$ indicates that the net benefit of coordination (i.e., $a-c$ or $b-c$) is always positive, whereas the net benefit of failing to cooperate (i.e., $-c$) is negative.
$-2\tilde{c}$ in the bottom row represents the \add{fundamental} cost of adopting the bilingual strategy $\sfab$. 
$\tilde{c}$ may be larger or less than $c$, depending on the extent to which the two memes are complementary or exclusive.
If $\tilde{c}\gg c$, then $\sfab$ will no longer be a plausible option since the two memes are prohibitively exclusive \add{(e.g., democrat/republican, Windows/Mac)}.
In contrast, when $\tilde{c}$ is low enough, $\sfab$ would be preferred to $\sfa$ and $\sfb$, because accepting a meme reduces the cost of accepting the other (e.g., MacBook and iPhone).

Neighbors' states are represented by vector ${\bf{m}}=(\mo,\ma,\mb,\mab)^\top$, where $m_{s}$ denotes the number of neighbors adopting strategy $s\in S$. Note that we have $\sum_{s\in S}m_s = k$ for nodes with degree $k$. 
The total payoff of a player having $k$ neighbors is given by the sum of the payoffs obtained by playing $k$ independent bilateral games~\cite{morris2000contagion,jackson2007diffusion,jackson-zenou2015games}.
We assume that the network has a locally tree-like structure, and that neighbors of a player are not directly connected.  
Therefore, in playing a game with a particular neighbor, the neighbor does not have an incentive to cooperate with other neighbors.
Let $v(s,\bf{m})$ denote the total payoffs of a player adopting strategy $s\in S$ and facing the neighbors' strategy profile $\vecm$. 
We have
\begin{align}
    v(\sfo,{\bf{m}}) &= 0, \label{eq:v_o}\\
    v(\sfa,{\bf{m}}) &= 
    -ck+aM_\sfa,   \label{eq:v_a} \\
    v(\sfb,{\bf{m}}) &= -ck + bM_\sfb,   \label{eq:v_b}\\
    v(\sfab,{\bf{m}}) &= -2\tilde{c}k + aM_\sfa+bM_\sfb, 
    \label{eq:v_ab}   
\end{align}
where $\Ma \equiv \ma+\mab$ (resp.~$\Mb \equiv \mb+\mab$) denotes the total number of neighbors accepting meme~$\sfa$ (resp.~$\sfb$), including bilinguals.
The optimal strategy $s^*$ is then expressed as a function of $\vecm$:
\begin{align}
    s^\ast(\vecm) = \underset{s\in S}{\rm arg\,max}\; v(s,\vecm).
    \label{eq:optimal_s}
\end{align}
In a time interval $dt$, a randomly chosen fraction $dt$ of $N$ individuals updates their strategies following Eq.~\eqref{eq:optimal_s}. 
It is assumed that the initial states are kept unchanged for nodes with $k=0$ since isolated nodes do not have a chance to play coordination game.

\subsection{Threshold rule as the optimal strategy in coordination games}

Based on the payoffs of each strategy \eqref{eq:v_o}--\eqref{eq:v_ab}, an individual optimally selects a strategy $s^\ast$ such that $v(s^\ast,\vecm)\geq v(s^\prime,\vecm)$ for all $s^\prime$: 
    
(i) $s^\ast =\sfa$ if $v_\sfa > v_\sfo$, $v_\sfa> v_\sfb$, and $v_\sfa> v_\sfab$:
\begin{align}
    -ck + a\Ma&>0, \\
    -ck + a\Ma &>-ck + b\Mb, \\
    -ck + a\Ma &>-2\tilde{c}k + a\Ma + b\Mb, 
\end{align}    
where $v_s$ is shorthand for $v(s,\vecm)$.
In the same manner, we have the following conditions for $s^\ast=\sfb$ and $\sfab$:

(ii) $s^\ast =\sfb$ if $v_\sfb > v_\sfo$, $v_\sfb> v_\sfa$, and $v_\sfb> v_\sfab$:
\begin{align}
    -ck + b\Mb&>0, \\
    -ck + b\Mb&>-ck + a\Ma, \\
    -ck + b\Mb&>-2\tilde{c}k + a\Ma + b\Mb, 
\end{align}    

(iii)  $s^\ast =\sfab$ if $v_\sfab > v_\sfo$, $v_\sfab> v_\sfa$, and $v_\sfab> v_\sfb$:
\begin{align}
    -2\tilde{c}k + a\Ma &+ b\Mb>0, \\
    -2\tilde{c}k + a\Ma &+ b\Mb  > -ck + a\Ma, \\
    -2\tilde{c}k + a\Ma &+ b\Mb > -ck + b\Mb.
\end{align}    
When there are ``tie'' strategies in simulation (i.e., $v_s = v_{s^\prime}$ for $s\neq s^\prime$), we randomly select a strategy among the tie strategies.

\begin{figure}[tb]
    \centering
      \includegraphics[width=7cm]{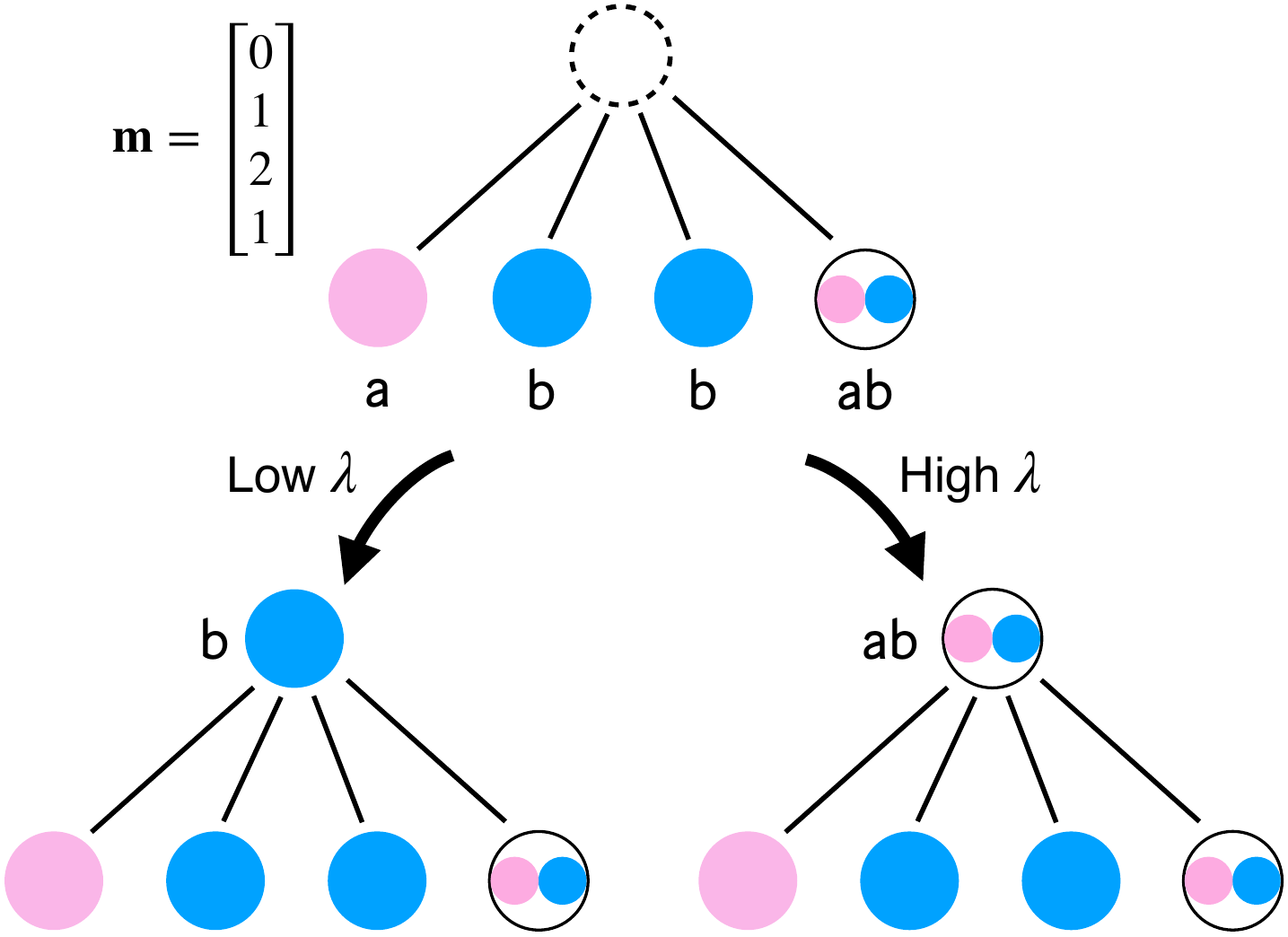}
    \caption{Schematic of strategic choice in the presence of multiple social memes.}
    \label{fig:schematic}
\end{figure}

\add{Given the above conditions}, the optimal strategy $s^\ast$ for each individual can be written as the following threshold rules:
\begin{equation} \label{eq:threshold}
s^\ast =
    \begin{cases}
    \sfa & \text{if }\; \frac{\Ma}{k}> \thetaa, \frac{\Mb}{k}<(1-\lambda)\thetab\; \text{ and }\frac{\Ma}{\Mb} > \frac{\thetaa}{\thetab}, \\
    \sfb &\text{if }\; \frac{\Mb}{k}> \thetab,\; 
    \frac{\Ma}{k}<(1-\lambda)\thetaa\; \text{ and }\; \frac{\Ma}{\Mb} < \frac{\thetaa}{\thetab}, \\
    \sfab & \text{if }\; \frac{\Ma}{k}> (1-\lambda)\thetaa,\; \frac{\Mb}{k}>(1-\lambda)\thetab\;  \\ 
    {} & \;\;\;\;\;\;\text{ and }\; \thetab\frac{\Ma}{k} +\thetaa\frac{\Mb}{k} > \thetaa \thetab (2-\lambda), \\
    \sfo & \text{otherwise}, 
   \end{cases}
 \end{equation}
where $\thetaa \equiv c/a\in(0,1)$, $\thetab \equiv c/b\in(0,1)$, and $\lambda \equiv 2(1-\widetilde{c}/c)$. $\lambda$ captures the degree of complementarity (or compatibility) between $\sfa$ and $\sfb$ where $\lambda >0$ (resp. $\lambda <0$) indicates that the two memes are complementary (resp. exclusive). When $\lambda=0$, they are mutually independent. 
In the analysis, we focus on a reasonable range of parameter values such that Nash equilibria of bilateral games are given by $(\sfo,\sfo)$, $(\sfa,\sfa)$, $(\sfb,\sfb)$ and $(\sfab,\sfab)$.
In fact, this assumption sets natural constraints for the threshold values: $\lambda <1$, $(1-\lambda)\thetaa<1$ and $(1-\lambda)\thetab<1$ (see Appendix~\ref{sec:constraint_lambda} for a derivation). 
Note that even if the neighborhood profile is the same, the optimal strategy may differ depending on $\lambda$ (Fig.~\ref{fig:schematic}).
If $\Mb=0$ (resp.\ $\Ma=0$), then the threshold rules reduce to the single threshold condition appeared in the binary-state cascade model \`{a} la Watts~\cite{Watts2002}: $m_\sfa/k > \thetaa$ (resp.\ $m_\sfb/k > \thetab$).

\subsection{Simulation procedure}

The procedure of numerical simulations is as follows:
\begin{enumerate}
     \item For given $z$ and $N$, generate an \ER network with a common connecting probability $z/(N-1)$. 
     \item Select seed nodes at random so that there are $\lfloor \rho^{\sfa}(0) N\rfloor$ nodes adopting strategy~$\sfa$ and $\lfloor \rho^{\sfb}(0) N\rfloor$ nodes adopting strategy~$\sfb$. The other nodes employ strategy~$\sfo$ as the status quo.
     \item Choose a fraction $dt\in (0,1)$ of nodes uniformly at random and update their strategies to maximize their payoffs $v$.
     \label{it:async}
     \item Repeat step 3 until convergence, where no nodes can be better off by changing their strategies.  
     \item Repeat steps 1--4.
 \end{enumerate}
Note that we implement an asynchronous update in step~\ref{it:async}, where a randomly chosen fraction $dt$ of nodes update their strategies in an infinitesimally small interval $dt$~\cite{gleeson2011high,gleeson2013binary}. 
We set $dt=0.01$ in all simulations.

\section{Results}

\subsection{AME solution}

In the present model, any of the three strategies $\{\sfa, \sfb,\sfab\}$ may spread globally, and the shares of each strategy in the stationary state, denoted by $\{\rho^s\}$, generally vary depending on the payoff parameters and network structure. This type of spreading process is considered as a multistate dynamical process, for which the \emph{approximate master equations} (AMEs) method has been used to analytically calculate the dynamical paths and the stationary state~\cite{gleeson2011high,gleeson2013binary,fennell2019multistate,kobayashi2021unstable} (see Appendix~\ref{sec:AME} for a description of the AME equations. The Matlab code is based on \cite{FennelCode}).

\begin{figure}[tb]
    \centering
      \includegraphics[width=8.6cm]{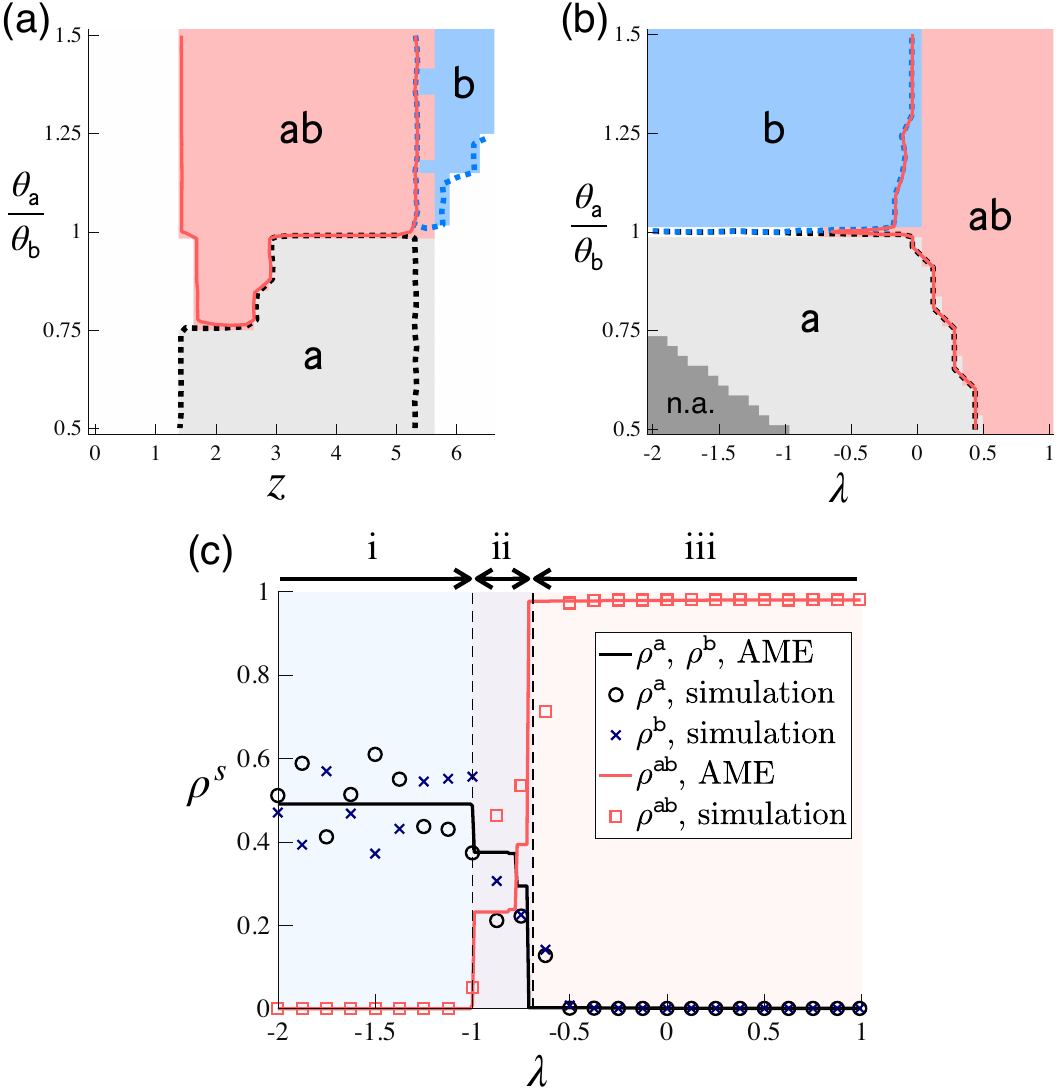}
    \caption{Phase diagram of equilibrium strategies. Relative threshold ($\thetaa/\thetab$) vs (a) mean degree $z$ ($\lambda=0$), and (b) complementarity $\lambda$ ($z=4$). Each colored area denotes a region within which a particular strategy is dominant (i.e., $\rho^s>0.5$) in the AME method based on \ER networks. Black dotted, blue dotted and red solid lines respectively indicate the boundaries of the dominant regions for $\sfa$, $\sfb$ and $\sfab$ obtained by simulation.  ``n.a.'' (shaded in dark gray) denotes the region in which the parameter constraints are not satisfied. 
    (c) Equilibrium share of each strategy obtained by simulation (symbols) and the AME method (lines). 
    There are three phases of social contagion, labeled as phases~i, ii and iii, depending on  $\lambda$.
    We set $a=4$, $b=4$ (in panel c), $c=1$ and $N = 10^4$. The average is taken over 100 runs with initial seed fraction $\rho^\sfa(0)=\rho^\sfb(0)=0.03$ and $\rho^\sfab(0)=0$.}
    \label{fig:phase_three_region}
\end{figure}

Depending on the inherent attractiveness (i.e., $a$ and $b$), the degree of complementarity $\lambda$ and the mean degree $z$, there are three phases as to which strategy is dominant in equilibrium (Figs.~\ref{fig:phase_three_region}a and b, and \ref{fig:shaded_SI}). 
We observe that the AME solutions (shaded) well predict the corresponding simulation results (lines).  
It should be noted that the \emph{cascade region}~\cite{Watts2002,Gleeson2007} within which we have $1-\rho^\sfo \gg 0$ is mostly covered by the combined dominant region (Fig.~\ref{fig:cascade_region_AME_sim}), suggesting that a strategy often dominates the others once a global cascade occurs.

While it is natural that the attractiveness parameters $a$ and $b$ explain the differences in popularity between $\sfa$ and $\sfb$ (Fig.~\ref{fig:phase_three_region}a and b), the following question still remains: 
what happens when the two memes are equally attractive (i.e., $a=b$) yet mutually exclusive?
When $a=b$,  we always have $\rho^{\sfa} = \rho^{\sfb}$ in the AME solution since there is no intrinsic difference between the two memes (black solid in Fig.~\ref{fig:phase_three_region}c).

\begin{figure*}[tb]
    \centering
    \includegraphics[width=13cm]{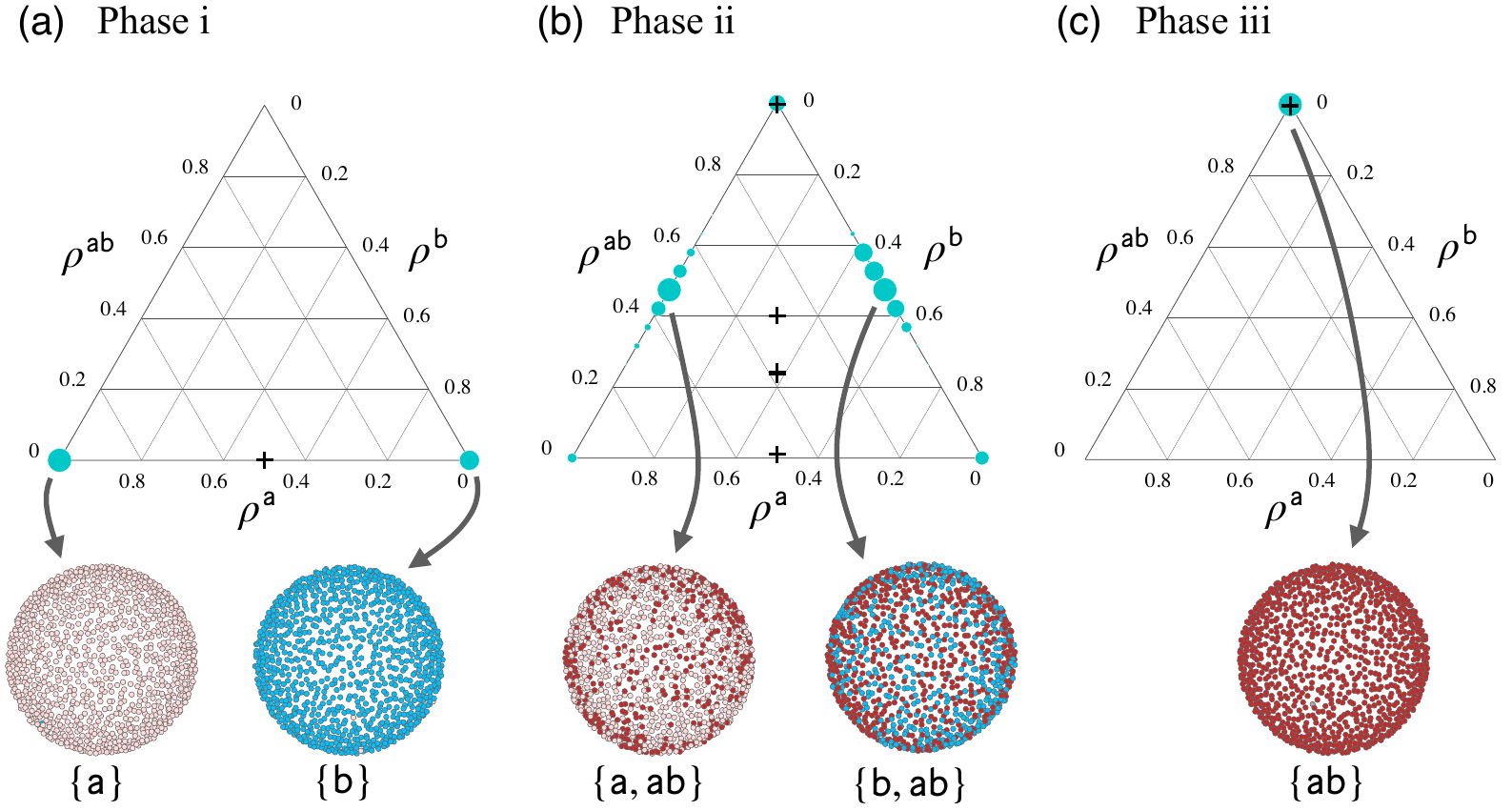}
    \caption{Ternary plot for the theoretical and simulated values of $(\rho^\sfa,\rho^\sfb,\rho^\sfab)$ in each phase.
    Phases~i--iii annotated at the top are defined based on the AME solution (see Fig.~\ref{fig:phase_three_region}c).
    The size of light-blue circle represents the simulated frequency, while black cross denotes the AME solution.
    We exclude simulation runs that did not reach convergence by $t= 300$.
    See the caption of Fig.~\ref{fig:phase_three_region} for the parameter values.}
    \label{fig:ternary_plot}
\end{figure*}

We find that there are three phases in the AME solutions \add{for the case of $\sfa =\sfb$}: \emph{i}) $\rho^\sfa,\rho^\sfb>0$ and $\rho^\sfab=0$, \emph{ii}) $\rho^\sfa,\rho^\sfb,\rho^\sfab>0$, and \emph{iii}) $\rho^\sfa=\rho^\sfb=0$ and $\rho^\sfab>0$ (Fig.~\ref{fig:phase_three_region}c).
\add{It is important to note that} while the stationary values of $\rho^\sfa$ and $\rho^\sfb$ are nearly $0.5$ in phase~i (i.e., $\lambda<-1$), this does not indicate that each of the strategies $\sfa$ and $\sfb$ is adopted by $50\%$ of the population. 
The average values for simulated $\rho^\sfa$ and $\rho^\sfb$ are nearly $0.5$ because the chance of $\sfa$ or $\sfb$ being a dominant strategy (i.e., $\rho^\sfa\approx 1$ or $\rho^\sfb\approx 1$) is close to $0.5$ (Fig.~\ref{fig:ternary_plot}a).
That is, the popularity of each meme is either 0\% or 100\% in each simulation (blue circle in the ternary plot in Fig.~\ref{fig:ternary_plot}a), although the fractions $\rho^\sfa$ and $\rho^\sfb$ averaged over simulation runs are both $0.5$, which corresponds to the AME value (black cross in Fig.~\ref{fig:ternary_plot}a).
This indicates that there is no diversity of memes (i.e., the two memes do not coexist) in a stationary state of a spreading process occurring in phase~i.

In phase~ii (i.e., $-1\lesssim\lambda\lesssim -0.7$), we have a different set of diffused strategies: $\{\sfa,\sfab\}$, $\{\sfb,\sfab\}$ and $\{\sfab\}$ (Fig.~\ref{fig:ternary_plot}b). 
The memes are neither too complementary nor too exclusive, and this is the only phase in which a strategy diversification may be observed. 
The AME solution indicates that $\rho^\sfa$ and $\rho^\sfb$ are less than $0.5$ (Fig.~\ref{fig:phase_three_region}c), but again strategies $\sfa$ and $\sfb$ do not coexist in simulation, resulting in a deviation from the theoretical average (Fig.~\ref{fig:ternary_plot}b).
In phase~iii (i.e., $\lambda > -0.7$), the two memes are not strongly mutually exclusive, so that the only strategy adopted in the stationary state is $\sfab$ (Fig.~\ref{fig:ternary_plot}, \emph{right}).
Since there is only one strategy that prevails in the network, the model is essentially the same as the binary-state cascade model, where the theoretical average is equal to the simulated popularity of $\sfab$ in each simulation.
These observations suggest that the intrinsic symmetry between the two types of memes leads to a symmetric cascade only in phase~iii while symmetry is likely to be broken in the other phases.


\begin{figure*}[tb]
    \centering
    \includegraphics[width=17cm]{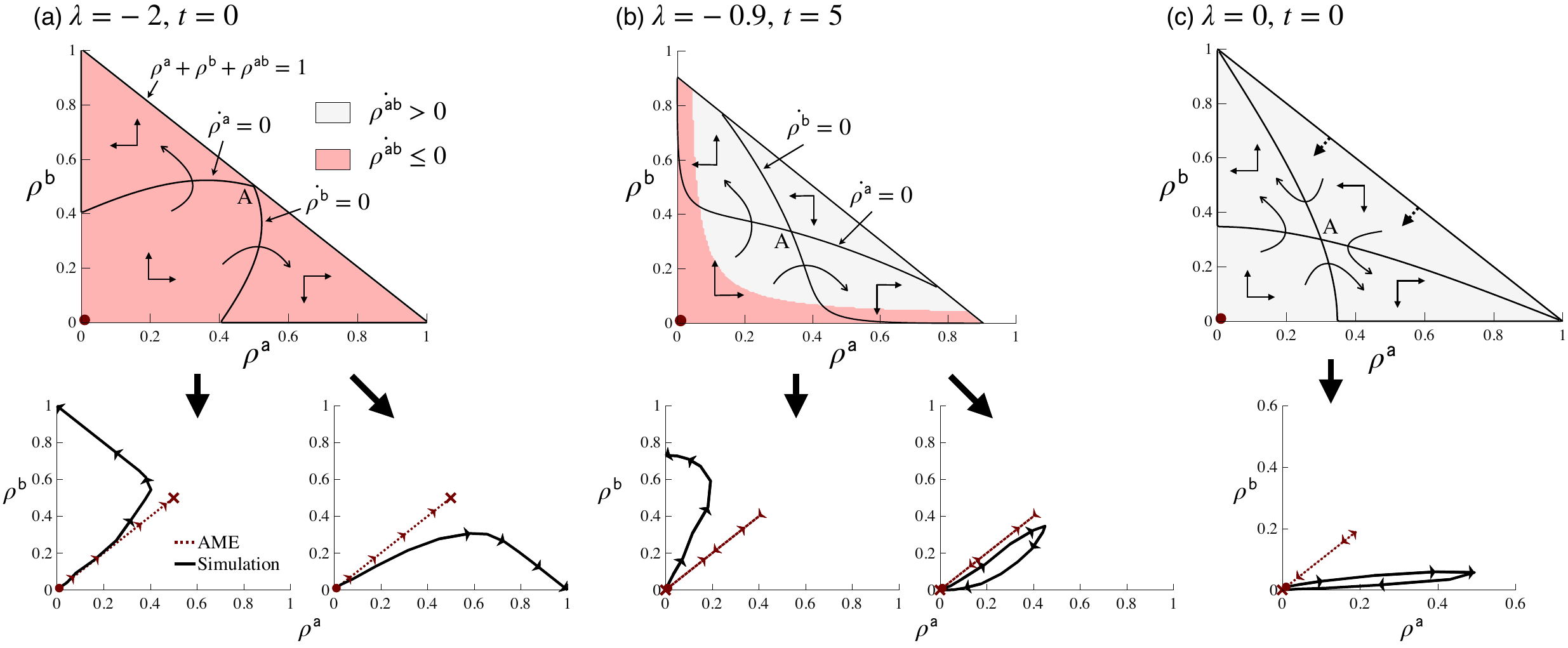}
    \caption{Phase diagram of the propagation process on 4-regular random graphs. 
    The initial state ($\rho^\sfa(0) = \rho^\sfb(0) = 0.01$, $\rho^\sfab(0)=0$) is indicated by red circle. 
    Point A is a saddle equilibrium, and examples of simulated path are shown at the bottom.
    The panels show typical behaviors in (a) phase~i, (b) phase~ii, and (c) phase~iii.
    We set $a=b=4$, $c=1$.}
    \label{fig:phase_diagram_MF}
\end{figure*}

\subsection{Mechanics of symmetry breaking}

To understand the fundamental mechanics behind the observed symmetry breaking, we draw phase diagrams based on a mean-field (MF) approximation using random $z$-regular networks (i.e., the degree distribution $p_{k}=\delta_{kz}$), for which it is assumed that the states of neighbors are independent of each other~\cite{bottcher2017critical}. 
In the MF method, the evolution of $\rho^s$ for each $s\in S$ is described by the following differential equation~\cite{gleeson2011high,gleeson2013binary,fennell2019multistate}:
\begin{align}
    \dot{\rho^s} =& -\sum_{s'\neq s}\rho^s\sum_{|\vecm|=z}\mathcal{M}_{z}(\vecm,\vecrho)F_\vecm(s\to s') \notag \\
     &+\sum_{s'\neq s}\rho^{s'}\sum_{|\vecm|=z}\mathcal{M}_{z}(\vecm,\vecrho)F_\vecm(s'\to s),
     \label{eq:MF_eq}
\end{align}
where $\vecrho \equiv (\rho^{\sfo},\rho^{\sfa},\rho^{\sfb},\rho^{\sfab})^\top$, and
$\mathcal{M}_z(\vecm,\vecrho)$ is the multinomial distribution given by
\begin{align}
    \mathcal{M}_z(\vecm,\vecrho) \equiv \frac{z!}{\mo!\ma!\mb!\mab!}(\rho^{\sfo})^{\mo}(\rho^{\sfa})^{\ma}(\rho^{\sfb})^{\mb}(\rho^{\sfab})^{\mab}.
\end{align}
$F_{\vecm}(s\to s^\prime)$ denotes the probability that individuals change their strategy from $s$ to $s^\prime$ for a given neighbors' profile $\vecm$: 
$F_{\vecm}(s\to s') =1$ if $s' = s^*(\vecm)$, and $0$ otherwise. 
The first term in Eq.~\eqref{eq:MF_eq} captures the rate at which a node changes its strategy from $s$ to $s^\prime(\neq s)$, and the second term denotes the rate at which a node newly employs strategy~$s$. Note that this is a system of four differential equations ($|S|=4$), but it is sufficient to use three of them because there is an obvious constraint $\sum_{s\in S} \rho^s=1$. 

Fig.~\ref{fig:phase_diagram_MF} presents phase diagrams in the $\rho^\sfa$-$\rho^\sfb$ space for three different values of $\lambda$, representing the phases~i--iii defined above.
Note that the theoretical equilibrium (indicated by point A) is saddle-path stable in all the three cases, but the diagrams differ in the size of the region in which $\dot{\rho^\sfab}>0$ (shaded in gray).
When the two memes are highly exclusive (Fig.~\ref{fig:phase_diagram_MF}a), there is no chance for strategy~$\sfab$ to gain popularity, so $\dot{\rho^{\sfab}}=0$ for any combination of $(\rho^\sfa,\rho^\sfb)$.
In simulations on finite-size networks, the saddle-path equilibrium indicated by the MF/AME method, $(\rho^\sfa,\rho^\sfb)=(0.5,0.5)$, is not practically reachable; simulated paths of $(\rho^\sfa,\rho^\sfb)$ converge to $(0,1)$ or $(1,0)$ once they deviate from the stable balanced path: $\rho^\sfa(t) = \rho^\sfb(t)$ for all $t\geq 0$ (red dotted in Fig.~\ref{fig:phase_diagram_MF}a, \emph{bottom}).

\begin{figure}[tb]
    \centering
    \includegraphics[width=8.3cm]{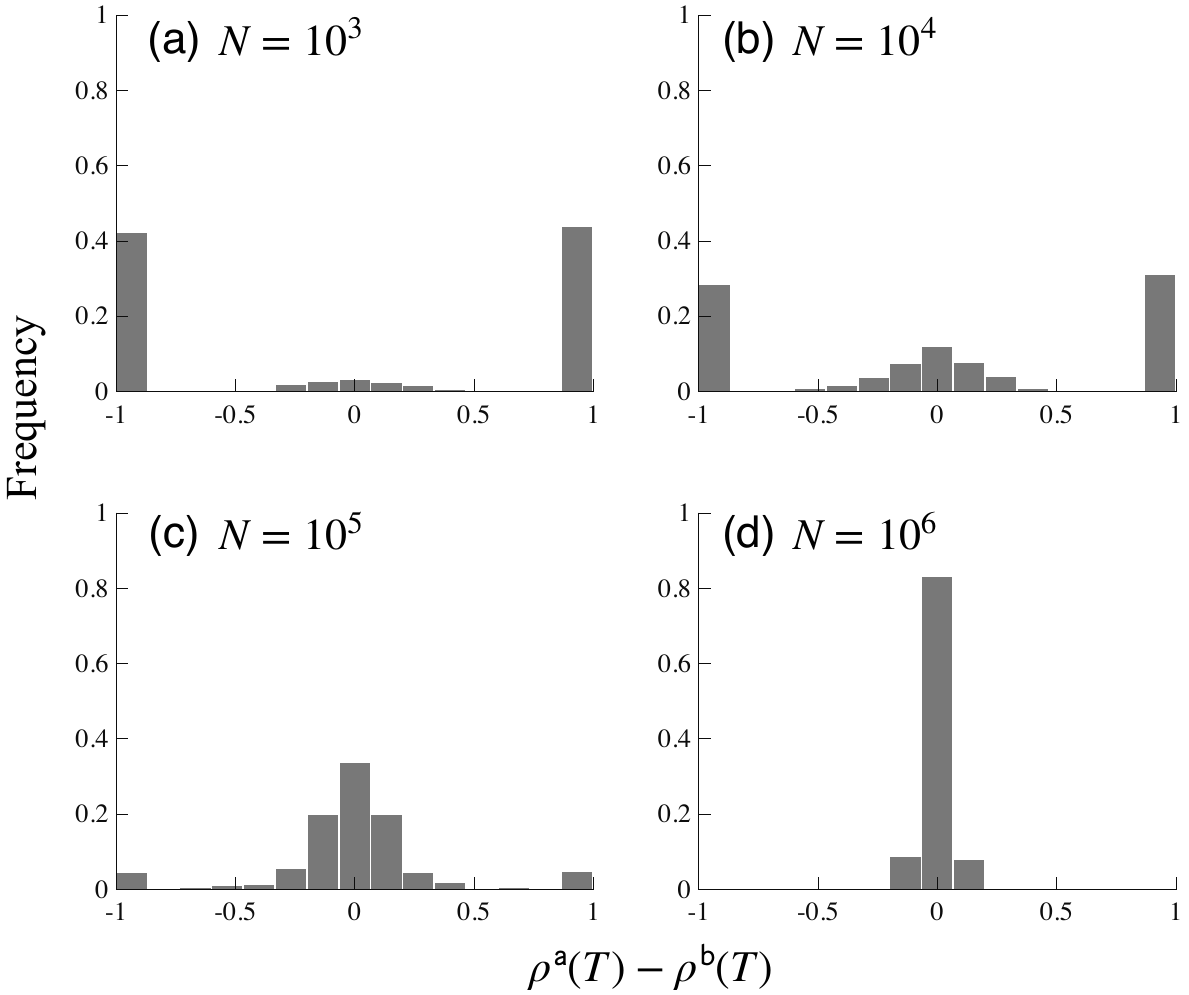}
    \caption{Size effect on the frequency of symmetry breaking.
     $a=b=1$, $c=1$, $\lambda=-1.5$, $z=4$, $\rho^\sfa(0)=\rho^\sfb(0)=0.1$, $\rho^\sfab(0)=0$, and $T=100$. We run 1,000 simulations for each network size.} 
    \label{fig:finite_size_effect}
\end{figure}

In principle, the symmetric MF/AME solution would correspond to the ``simulated'' equilibrium in the limit of large networks with no structural fluctuations.
However, any \add{synthetically generated} networks are generally not free from finite-size effects and fluctuations, so it is not guaranteed that $\rho^\sfa(t) = \rho^\sfb(t)$ for all $t\geq 0$ \add{in simulations}. 
Histograms of simulated values of $\rho^\sfa-\rho^\sfb$ at a certain point in time, denoted by $T$, reveal the effect of network size on the likelihood of symmetry breaking (Fig.~\ref{fig:finite_size_effect}).
When $N$ is relatively small, symmetry breaking occurs in the early stage of spreading process, so we often have $\rho^\sfa(T)=1$ or $\rho^\sfb(T)=1$ at $T=100$ (Fig.~\ref{fig:finite_size_effect}a and b).
In contrast, when $N=10^5$ or larger, it is much less likely that either of the strategies is adopted by most of the population at $T=100$, indicating that the intrinsic symmetry of the memes is more likely to be maintained for larger networks (Fig.~\ref{fig:finite_size_effect}c and d).

In phase~ii, there arises an area in which $\dot{\rho^{\sfab}}>0$ (Fig.~\ref{fig:phase_diagram_MF}b).
This suggests that the feasible region of $(\rho^\sfa,\rho^\sfb)$ (i.e., $\{(\rho^\sfa,\rho^\sfb): \rho^\sfa \geq 0, \rho^\sfb \geq 0, \rho^\sfa +\rho^\sfb+\rho^\sfab\leq 1 \}$) gradually shrinks as $\rho^\sfab$ increases as long as the current state of $(\rho^\sfa,\rho^\sfb)$ is in the gray-shaded area.
In this phase, symmetry breaking may occur, but not always (Fig.~\ref{fig:phase_diagram_MF}b, \emph{bottom}).
In the latter case, both $\rho^\sfa$ and $\rho^\sfb$ initially increase and then begin to decrease as the feasible region shrinks in accordance with a rise in $\rho^\sfab$.
In phase~iii, we always have $\dot{\rho^{\sfab}}>0$ (Fig.~\ref{fig:phase_diagram_MF}c).
This indicates that any path of $(\rho^\sfa,\rho^\sfb)$ will move toward the origin at some point in time as $\rho^\sfab$ increases.
Therefore, $\sfab$ will be the only diffused strategy in equilibrium.
The time to reach convergence in simulated cascades follows a heavy-tailed distribution when symmetry breaking always occurs (i.e., in phase~i), while the spreading process promptly reaches an equilibrium in the other phases (Fig.~\ref{fig:ccdf_convtime}).

\subsection{Irreversibility of individual behavior}

In the model shown above, individuals' choices are fully \emph{reversible} where the past strategies do not affect the current strategic choice (Eq.~\ref{eq:optimal_s}). This is a reason why either of the social memes could dominate the other and there is no possibility of \emph{polarization}: $\rho^\sfa\gg 0,\; \rho^\sfb\gg 0$ and $\rho^\sfab = 0$~\cite{vasconcelos2019consensus,baumann2020modeling}.
Such a reversible decision making, however, would be practically infeasible when switching costs are high (e.g., switching from Mac to Windows).
To investigate irreversible dynamics, we introduce a parameter $q\in [0,1]$ representing the degree of irreversibility; $q=0$ and $1$ respectively correspond to the fully reversible and irreversible cases.
When $q=1$, only the following five switching patterns are allowed: $\sfo\to\sfa$, $\sfo\to \sfb$, $\sfo\to \sfab$, $\sfa\to\sfab$, and $\sfb\to\sfab$. 
Thus, once a meme is accepted, there is no possibility that the meme will be abandoned (i.e., $\sfa\nrightarrow\sfo$, $\sfa\nrightarrow\sfb$, $\sfab\nrightarrow\sfb$, etc).
The irreversibility parameter $q\in [0,1]$ denotes the rate at which a strategy will not be reverted. 
The response function with irreversibility constraints, denoted by $\widetilde{F}_\vecm (s\to s^\prime)$, is given in Table~\ref{tab:response_irreversible}:

\begin{table*}[tbh]
    \centering
    \caption{Elements of the irreversible response function $\widetilde{F}_\vecm(s\to s^\prime)$.
    The unconstrained response function $F_\vecm$ is defined by Eq.~\eqref{eq:response_func}. 
    Rows and columns denote the current states (i.e., $s$) and the next states (i.e., $s^\prime$), respectively. 
    }
    \begin{tabular}{llcccc}
         & & &\multicolumn{2}{c}{$s^\prime$} & \\
             \hline
          &    & $\sfo$ & $\sfa$ & $\sfb$ & $\sfab$ \\
              \hline
    &    $\sfo\;\;\;$ &  $F_\vecm(\sfo\to\sfo)$ & $F_\vecm(\sfo\to\sfa)$ &$F_\vecm(\sfo\to\sfb)$ & $F_\vecm(\sfo\to\sfab)$ \\
     \multirow{2}*{$s\;\;$} &   $\sfa\;\;\;$ & $(1-q)F_\vecm(\sfa\to\sfo)$ & $1-\sum_{s\neq \sfa}\widetilde{F}_\vecm(\sfa\to s)$ & $(1-q)F_\vecm(\sfa\to\sfb)$ & $F_\vecm(\sfa\to\sfab)$ \\
             &
        $\sfb\;\;\;$ & $(1-q)F_\vecm(\sfb\to\sfo)$ & $(1-q)F_\vecm(\sfb\to\sfa)$ & $1-\sum_{s\neq \sfb}\widetilde{F}_\vecm(\sfb\to s)$ & $F_\vecm(\sfb\to\sfab)$ \\
     &   $\sfab\;\;\;$ & $(1-q)F_\vecm(\sfab\to\sfo)$ & $(1-q)F_\vecm(\sfab\to\sfa)$ & $(1-q)F_\vecm(\sfab\to\sfb)$ & $1-\sum_{s\neq \sfab}\widetilde{F}_\vecm(\sfab\to s)$ \\
        \hline
    \end{tabular}
    \label{tab:response_irreversible}
\end{table*}
 For nodes with $s=\sfo$, there is no constraint in updating their strategy.
 For nodes with $s=\sfa$ (resp.\ $s=\sfb$), shifting to $s^\prime=\sfb$ (resp.\ $s^\prime =\sfa$) or $s^\prime=\sfo$ is restricted, for which the transition probability is multiplied by a factor of $(1-q)$. 
 For nodes with $s=\sfab$, any state change is restricted.
 Note that the unconstrained response function is recovered if $q=0$.

\begin{figure}[tb]
    \centering
    \includegraphics[width=8.6cm]{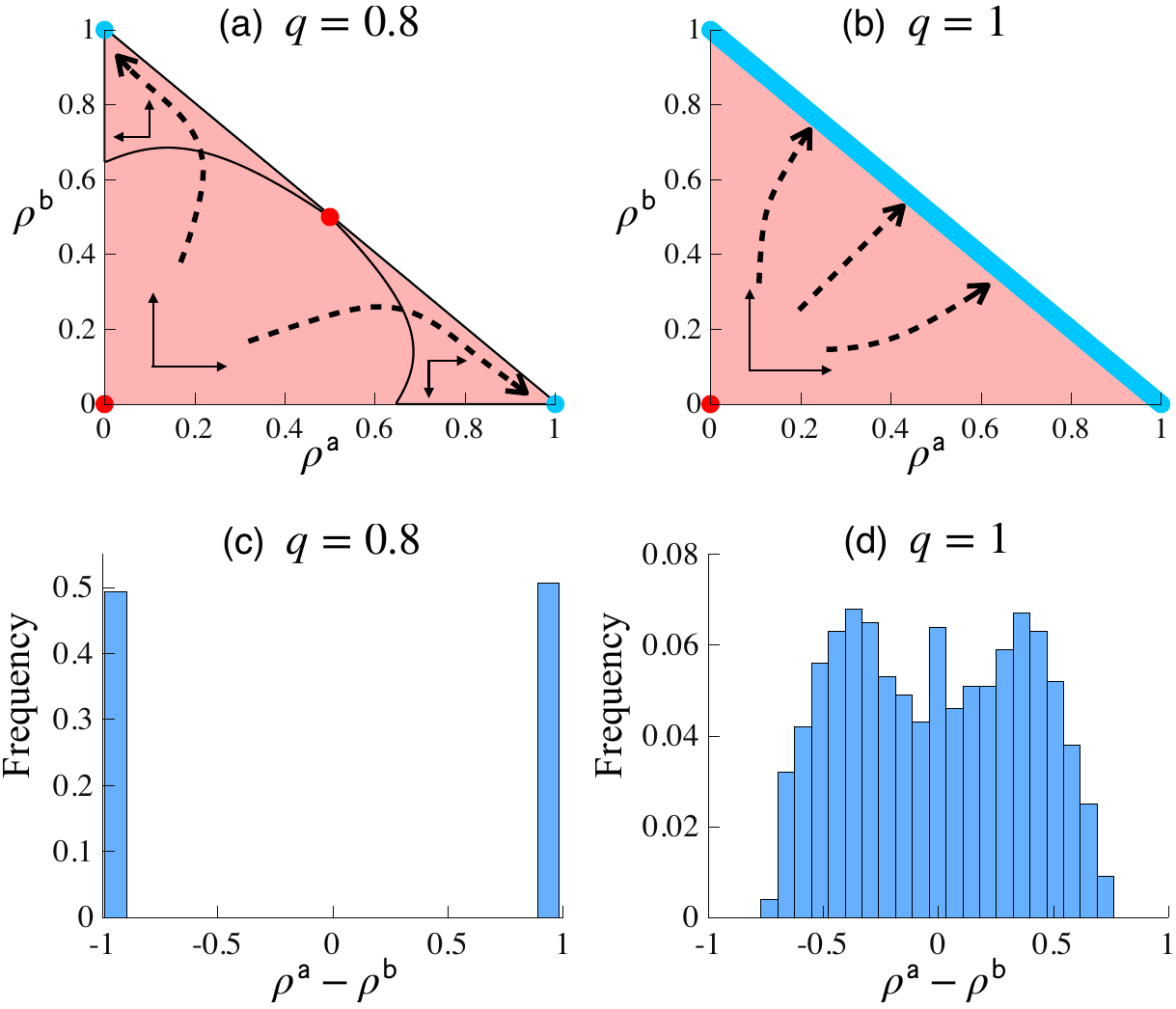}
    \caption{Equilibrium indeterminacy due to irreversibility. Phase diagrams for (a) partially irreversible ($q=0.8$), and (b) fully irreversible ($q=1$) cases. Histograms of $\rho^\sfa-\rho^\sfb$ for (c) $q=0.8$ and (d) $q=1$.  
    The phase diagrams are obtained using the MF method based on 4-regular random graphs, while the histograms \add{at the stationary state} are obtained from simulations using \ER graphs with $z=4$. 
    In panels a and b, blue and red circles respectively denote stable and unstable equilibria at which $\dot{\rho^s}=0, \forall s\in S$.
    We run simulations 1,000 times with $\rho^\sfa(0)=\rho^\sfb(0)=0.01$, $\rho^\sfab(0)=0$,
    $a=b=4$, $c=1$, and $\lambda=-2$ in all panels. 
    }
    \label{fig:phase_maxeig}
\end{figure}

Let $G_s$ be a function that represents the right-hand side of the MF equation~\eqref{eq:MF_eq} (i.e., $\dot{\rho}^s=G_s(\vecrho)$).
A stable (resp.\ unstable) equilibrium is defined as an equilibrium at which $\dot{\rho}^s=0$ for all $s\in S$ and the maximum eigenvalue of the Jacobian of vector $\mathbf{G} = (G_\sfo,G_\sfa,G_\sfb,G_\sfab)^\top$ is non-positive (resp.\ positive).
We find that introducing a partial irreversibility (i.e., $q<1$) does not qualitatively change the dynamical process; there are still two symmetric unstable equilibria,  $(\rho^\sfa,\rho^\sfb)=(0,0)$ and $(0.5,0.5)$ (red circles in Fig.~\ref{fig:phase_maxeig}a), and two asymmetric stable equilibria, $(0,1)$ and $(1,0)$ (blue circles in 
Fig.~\ref{fig:phase_maxeig}a).
Symmetry breaking always occurs in phase~i as in the fully reversible model (Fig.~\ref{fig:phase_maxeig}c).
Note, however, that the greater the degree of irreversibility $q$, the longer the time to convergence for $q<1$ (Fig.~\ref{fig:convtime_vs_q}).

In phase~i, where $\rho^\sfab(t)=0$ for all $t$, the saddle equilibrium disappears when the strategies are fully irreversible (i.e., $q=1$).
Instead, there arises a continuum of stable equilibria $({\rho}^\sfa,{\rho}^\sfb)$ such that ${\rho}^\sfa+{\rho}^\sfb = 1$ (Fig.~\ref{fig:phase_maxeig}b).
This indicates that equilibrium is indeterminate in irreversible dynamics even in the limit of large networks.
Indeed, the simulated equilibria are continuously distributed, at each of which polarization occurs (i.e., $\rho^\sfa\gg 0,\; \rho^\sfb\gg 0$ and $\rho^\sfab =  0$) (Fig.~\ref{fig:phase_maxeig}d). 
This is intuitive given that the state-transition process is no longer ergodic when $q=1$.
Due to the irreversibility, the time to convergence is minimized at $q=1$ (Fig.~\ref{fig:convtime_vs_q}).
We also find that in phases~ii and iii, the bilingual strategy $\sfab$ promptly becomes the dominant strategy when $q=1$ (Figs.~\ref{fig:irreversible_lambda-09} and \ref{fig:irreversible_lambda0}).

\section{Discussion}

 We presented a generalized model of complex social contagion with multiple social memes based on a game-theoretic foundation. 
 The model explains how symmetry breaking and polarization occur in the spread of competing information on networks.
 While the ``average'' popularity of each meme can be well approximated by the AME/MF methods, averaging is not appropriate when symmetry is broken in the actual spreading process.

 There are some issues to be addressed in future research.
 First, the proposed model based on coordination games should be regarded as an example of possible extensions of the cascade model for which individual behavior is rationalized. 
 While the current work provides a microfoundation of the Watts threshold model from a game-theoretic approach, different specifications of strategic behavior could lead to different forms of threshold rules.
 
Second, we did not consider any non-random network structure, such as community structure.
 The absence of community structure might be a reason why polarization does not occur in the case of reversible strategies. 
 Third, unlike the binary-state cascade models, it is difficult to obtain analytical conditions under which a global cascade can occur. 
 We exploited the power of AMEs to show the boundary of cascade region, yet a simple analytical cascade condition would be useful to predict global cascades.

T.\ K.\ acknowledges  financial support from JSPS KAKENHI\ 19H01506 and 20H05633. I would like to thank Tomokatsu Onaga for useful comments.

\appendix

\section{Constraints for $\lambda$}\label{sec:constraint_lambda}

Since we focus on a situation in which the pure strategy Nash equilibria for each bilateral game are given by $(\sfo,\sfo)$, $(\sfa,\sfa)$, $(\sfb,\sfb)$ and $(\sfab,\sfab)$, the payoff of strategy~$s$ must be the highest if the opponent's strategy is $s$. 
We have the following conditions for each of these strategy pairs to be attained as a Nash equilibrium:
\begin{enumerate}
    \item[(i)] For the strategy pair $(\sfo,\sfo)$ to be a Nash equilibrium, we need to have $-2\tilde{c}<0$. 
    Since $\lambda =2(1-\tilde{c}/c)$, it indicates that 
    \begin{align}
    \lambda<2.
    \label{eq:lambda_sfo}
    \end{align}
    \item[(ii)] For the strategy pair $(\sfa,\sfa)$ to be a Nash equilibrium, we need to have $a-c>a-2\tilde{c}$. It follows that
    \begin{align}
    \lambda<1.
    \label{eq:lambda_sfa}
    \end{align}
    Note that the condition for the pair $(\sfb,\sfb)$ is the same.
    \item[(iii)] For the strategy pair $(\sfab,\sfab)$ to be a Nash equilibrium, we need to have $a+b-2\tilde{c}>a-c$ and $a+b-2\tilde{c}>b-c$ (Recall that $a-c>0$ and $b-c>0$). It follows that
    \begin{align}
        (1-\lambda)\thetaa < 1 \;\;\text{ and }\;\; (1-\lambda)\thetab < 1.
        \label{eq:lambda_sfab}
    \end{align}
\end{enumerate}
Given the conditions \eqref{eq:lambda_sfo}--\eqref{eq:lambda_sfab}, $\lambda$ must satisfy $\lambda<1$, $(1-\lambda)\thetaa < 1$, and $(1-\lambda)\thetab < 1$.

\section{AME equations}\label{sec:AME}

Here, we describe the spreading process of competing memes based on the AME method. 
Let $\rho^{s}_{k,\vecm}$ denote the fraction of $k$-degree nodes belonging to the $(s,\vecm)$ class (i.e., $k$-degree nodes adopting strategy~$s$ and facing the neighbor profile $\vecm$). 
Using the AME formalism, the evolution of $\rho^{s}_{k,\vecm}$ is given by~\cite{gleeson2011high,gleeson2013binary,fennell2019multistate}:
\begin{align}
     \dot{\rho^{s}}_{k,\vecm} \;=\; & -\sum_{s' \neq s}F_{\vecm}(s\to s')\rho_{k,\vecm}^{s} \notag \\
     & -\sum_{r\in S}\sum_{r'\neq r}m_{r}\phi_{s}(r\to r')\rho_{k,\vecm}^{s} \notag \\
     & +\;\: \sum_{s' \neq s}F_{\vecm}(s'\to s)\rho_{k,\vecm}^{s'}
    \notag \\ &+  \sum_{r\in S}\sum_{r'\neq r}(m_{r'}+1)\phi_{s}(r'\to r)\rho_{k,\vecm-{\bf e}_{r} + {\bf e}_{r'}}^{s},
\label{eq:AME_eqs}
\end{align}
for $s\in S$, where $\phi_s(r\to r^\prime)$ denotes the probability that a neighbor of a node adopting strategy~$s$ changes its strategy from $r$ to $r^\prime$: 
\begin{align}
    \phi_s(r\to r') =& \frac{\sum_k p_k \sum_{|\vecm|=k} m_s{\rho}^{r}_{k,\vecm}F_\vecm(r\to r')}{\sum_k p_k \sum_{|\vecm|=k} m_s \rho_{k,\vecm}^{r}}. \label{eq:phi}
\end{align}
$p_k$ denotes the degree distribution, and the response function $F_{\vecm}(s\to s^\prime)$ describes the rate at which individuals change their strategy from $s$ to $s^\prime$ for a given neighbors' profile $\vecm$:
\begin{align}
    F_{\vecm}(s\to s') =
    \begin{cases}
    1  \;\;{\text{ if }}\; s' = s^*(\vecm),  \\ 
    0 \;\;{\text{ otherwise}},
    \end{cases}
    \label{eq:response_func}
\end{align}
where $s^*(\vecm)$ is the optimal strategy defined in Eq.~\eqref{eq:optimal_s}. The expected fraction of individuals adopting strategy $s\in S$ leads to $\rho^s = \sum_k p_k\sum_{|\vecm|=k} \rho_{k,\vecm}^s$, where $\sum_{|\vecm|=k}$ denotes the sum over all combinations of $\{m_s\}$ such that $\sum_{s\in S}m_s=k$. 

There are four factors that change $\rho_{k,\vecm}^s$ over time in Eq.~\eqref{eq:AME_eqs}.  
Individuals will \emph{leave} the $(s,\vecm)$ class if \emph{i}) their strategy changes from $s$ to $s'(\neq s)$ (the first term) or \emph{ii}) their neighbor profile changes from $\vecm$ to $\vecm' (\neq \vecm)$ (the second term). 
Individuals will \emph{enter} the $(s,\vecm)$ class if \emph{iii}) their strategies newly change from $s'(\neq s)$ to $s$ (the third term) or \emph{iv}) the neighbors' profile shifts from $\vecm'(\neq \vecm)$ to $\vecm$ (the fourth term).
The expression $\vecm-{\bf e}_r+{\bf e}_{r'}$ in the fourth term denotes the neighbor profile that has $m_{r'}+1$ in the $r'$-th element and $m_{r}-1$ in the $r$-th element.

The denominator of Eq.~\eqref{eq:phi}, $\sum_k p_k \sum_{|\vecm|=k} m_s\rho_{k,\vecm}^r$, represents the expected number of $(s)$--$(r)$ edges. 
Since the expected number of $(s)$--$(r)$ edges that shift to $(s)$--$(r')$ in an infinitesimal interval $dt$ is given as $\sum_k p_k \sum_{|\vecm|=k} m_s\rho_{k,\vecm}^rF_\vecm(r\to r')dt$, the probability of a $(s)$--$(r)$ edge shifting to a $(s)$--$(r')$ edge, denoted by $\phi_s(r\to r')dt$, is obtained as the ratio of the two, leading to Eq.~\eqref{eq:phi}.
The AME solution is calculated using Matlab codes provided in~\cite{FennelCode}.


%

\newpage
\clearpage
\clearpage

\setcounter{section}{0}
\setcounter{table}{0}
\setcounter{equation}{0}
\setcounter{figure}{0}
\setcounter{page}{1}
     
\renewcommand{\thetable}{S\arabic{table}}
\renewcommand{\thefigure}{S\arabic{figure}}
\renewcommand{\thesection}{S\arabic{section}}
\renewcommand{\theequation}{S\arabic{equation}}
\begin{widetext}

{\flushleft
{\fontsize{16pt}{16pt}\selectfont
 \textbf{Supplemental Material} \\
 \vspace{.7cm}
 \Large{``\papertitle''} \\
 \vspace{.5cm}
{\large Teruyoshi Kobayashi}
}
}
\vspace{2cm}

\begin{figure}[thb]
    \centering
    \includegraphics[width=12cm]{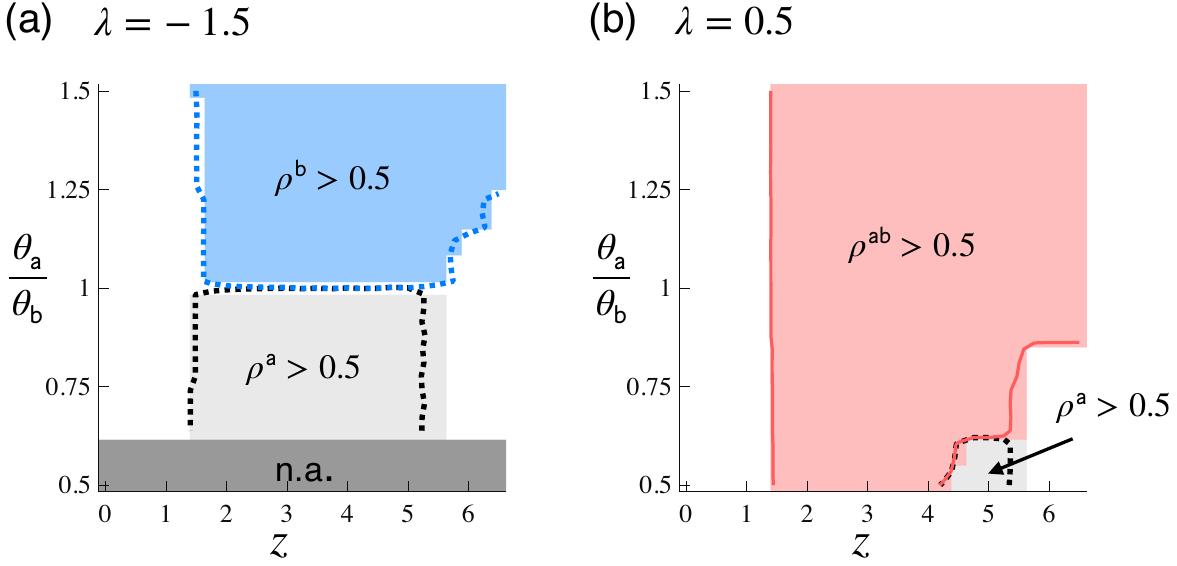}
    \caption{Dominant strategy in the stationary state.
    See the caption of Fig.~\ref{fig:phase_three_region} for details.}
    \label{fig:shaded_SI}
\end{figure}

\vspace{1.5cm}

\begin{figure}[thb]
    \centering
    \includegraphics[width=15.5cm]{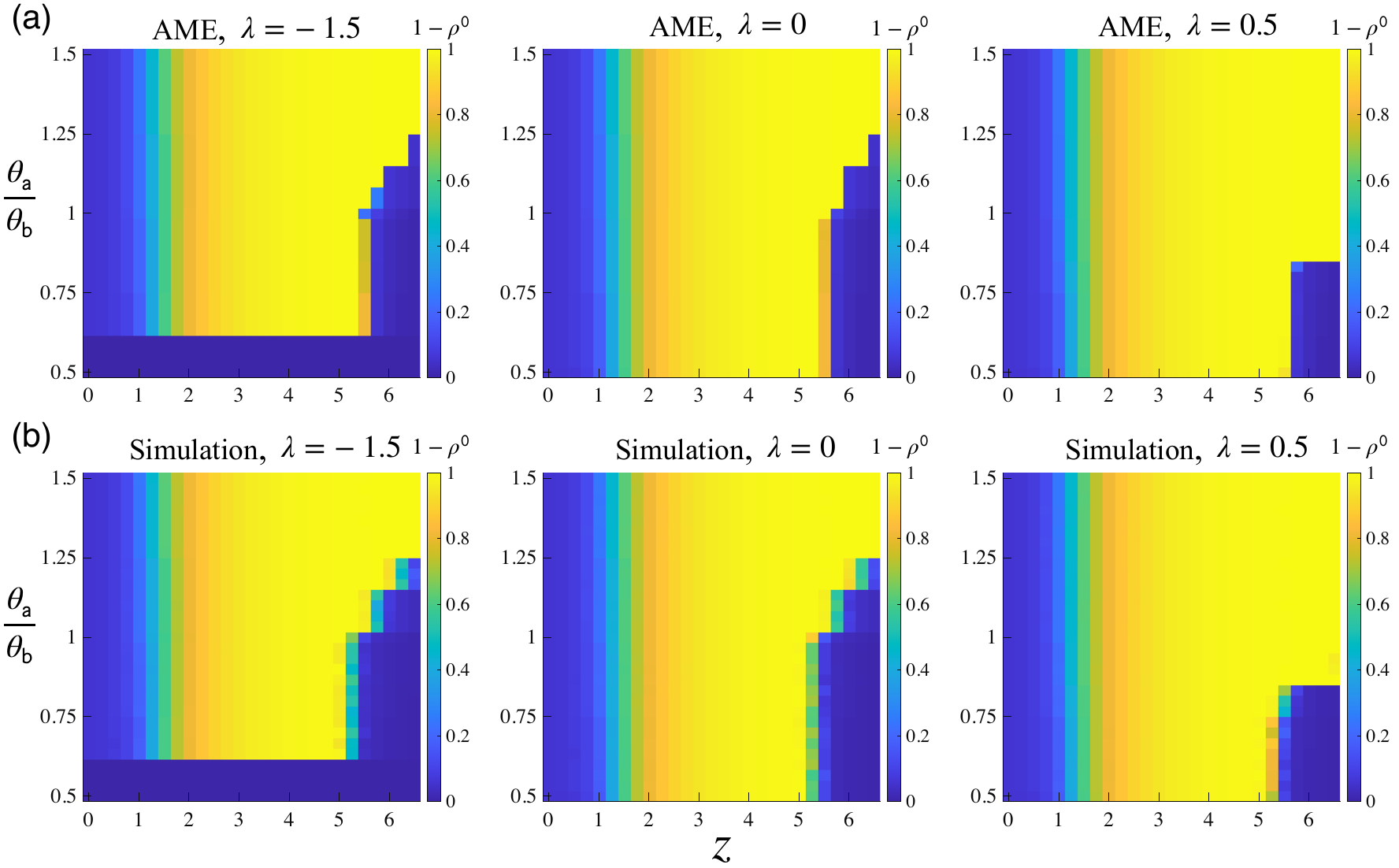}
    \caption{Theoretical and simulated cascade region. (a) AME solution and (b) simulation. 
    Color indicates the value of $\rho^\sfa + \rho^\sfb +\rho^\sfab$ $(=1-\rho^\sfo)$, which will be significantly larger than $\rho^\sfa(0) + \rho^\sfb(0) +\rho^\sfab(0)= 0.06$ if a global cascade occurs.
    See the caption of Fig.~\ref{fig:phase_three_region} for the parameter values.}
    \label{fig:cascade_region_AME_sim}
\end{figure}
\clearpage

\vspace{1.5cm}

\begin{figure}[thb]
    \centering
    \includegraphics[width=16cm]{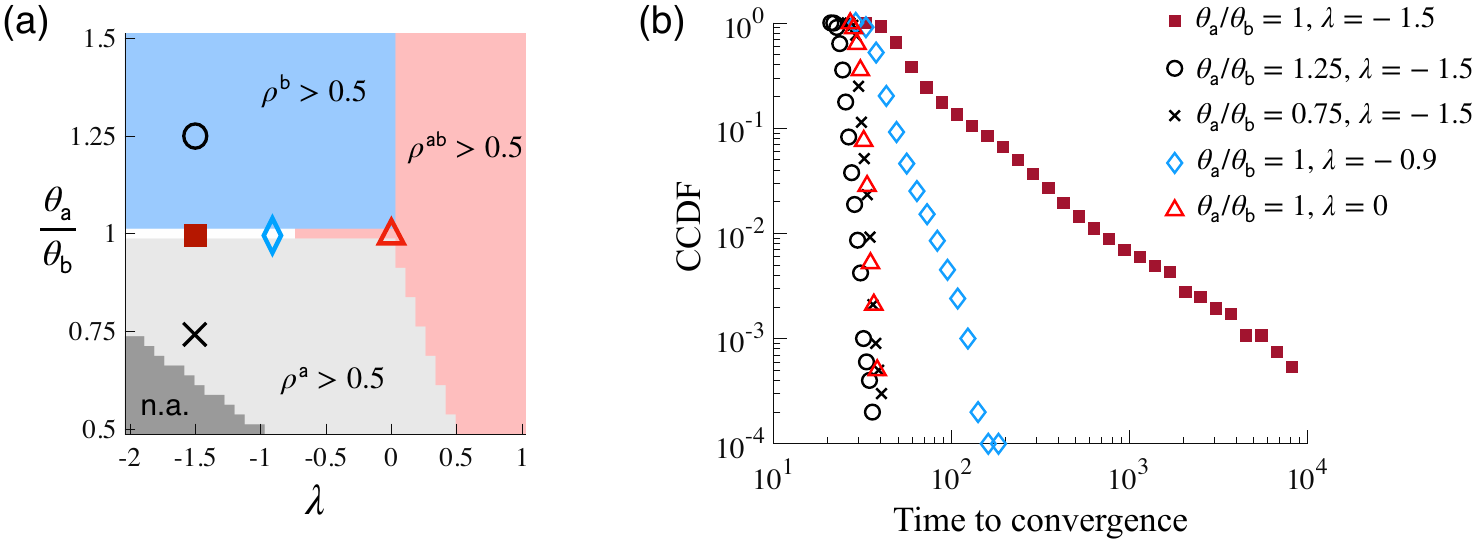}
    \caption{Distribution of the time to convergence. (a) Dominant regions as in Fig.~\ref{fig:phase_three_region}b. A symbol denotes a point at which a complementary cumulative distribution function (CCDF) of convergence time is generated in panel~b.
    (b) CCDF of the time to convergence. Each symbol denotes a particular parameter combination indicated in panel~a. When $\thetaa/\thetab=1$ and $\lambda=-1.5$ (i.e., in phase~i), at which symmetry breaking always occurs, simulated time to convergence follows a heavy-tailed distribution. We conduct $10,000$ simulations on \ER networks with $z=4$ and discard simulation runs that did not reach convergence by $t=10,000$.
    See the caption of Fig.~\ref{fig:phase_three_region} for the other parameter values.
    }
    \label{fig:ccdf_convtime}
\end{figure}

\vspace{1.5cm}

\begin{figure}[thb]
    \centering
    \includegraphics[width=8.5cm]{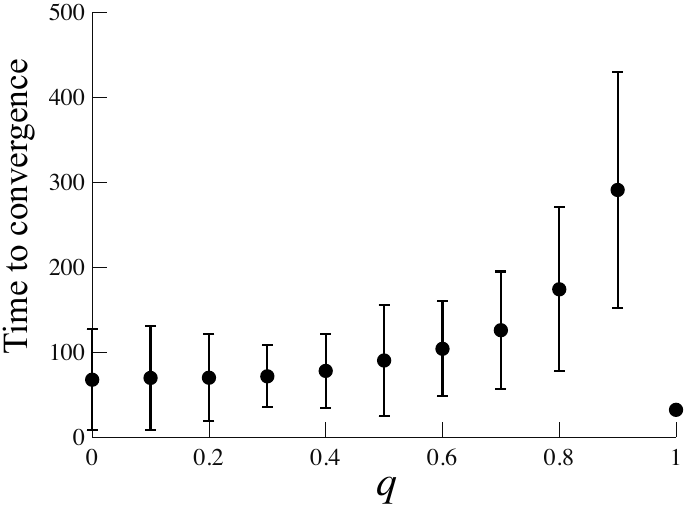}
    \caption{Time to convergence and the degree of irreversibility $q$. 
    Error bar denotes one standard deviation while circle denotes the average.
    See the caption of Fig.~\ref{fig:phase_maxeig} for the details of simulation. }
    \label{fig:convtime_vs_q}
\end{figure}

\vspace{1.5cm}

\begin{figure}[thb]
    \centering
    \includegraphics[width=16.5cm]{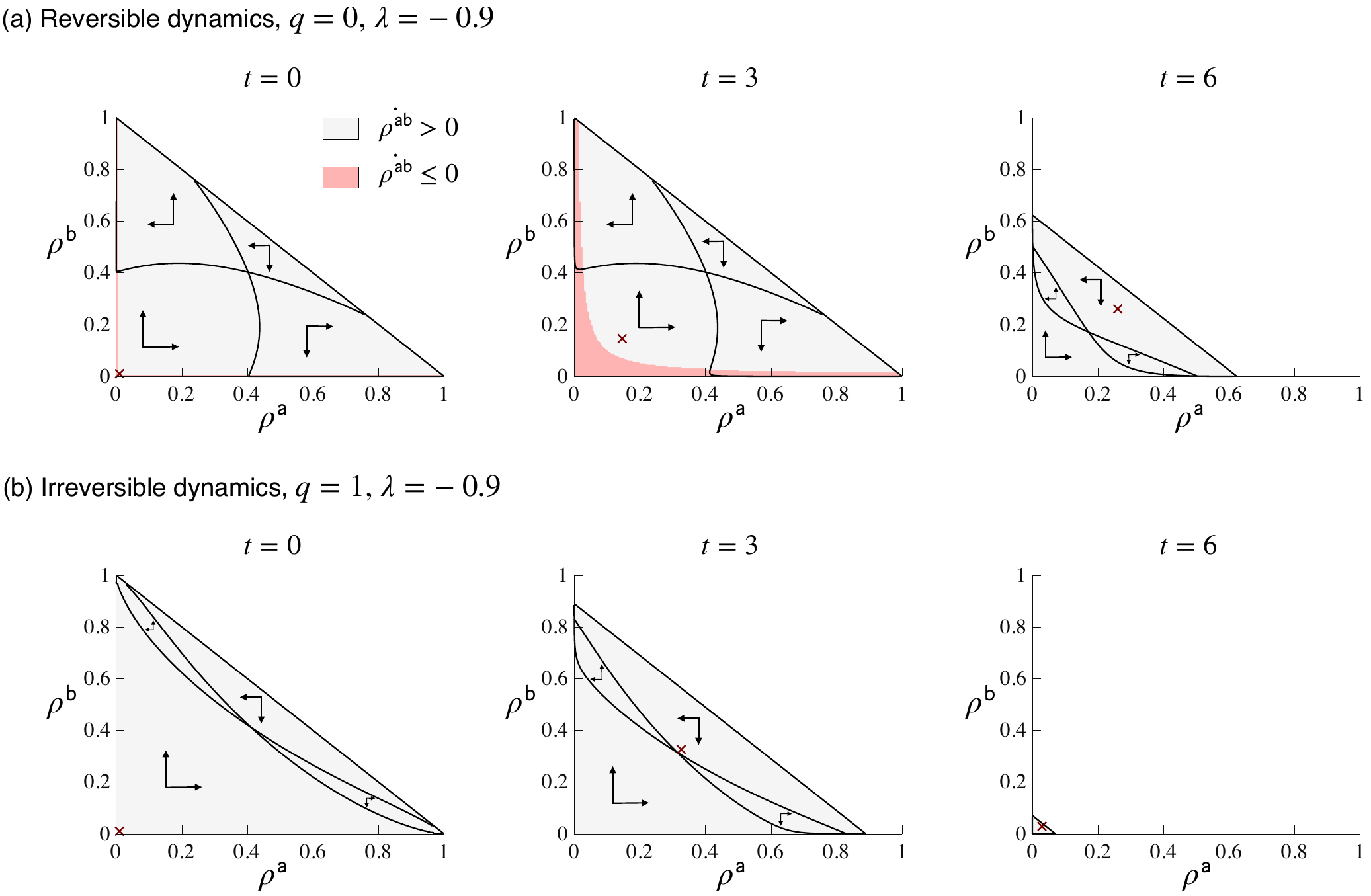}
    \caption{Phase diagrams of (a) reversible and (b) irreversible dynamics for $\lambda=-0.9$.
    See Fig.~\ref{fig:phase_diagram_MF} for a detailed description of the phase diagrams.
    $q$ denotes the extent to which a strategy is irreversible (i.e., $q=0$ and $1$ represent fully reversible and irreversible cases, respectively).
    Red cross denotes the MF solution.
    In panel (b), due to the presence of irreversibility constraints, the popularity of $\sfab$ increases faster than in the case of $q=0$, which shrinks the feasible region of $(\rho^\sfa,\rho^\sfb)$ faster along with it.
    }
    \label{fig:irreversible_lambda-09}
\end{figure}

\vspace{1.5cm}

\begin{figure}[thb]
    \centering
    \includegraphics[width=16.5cm]{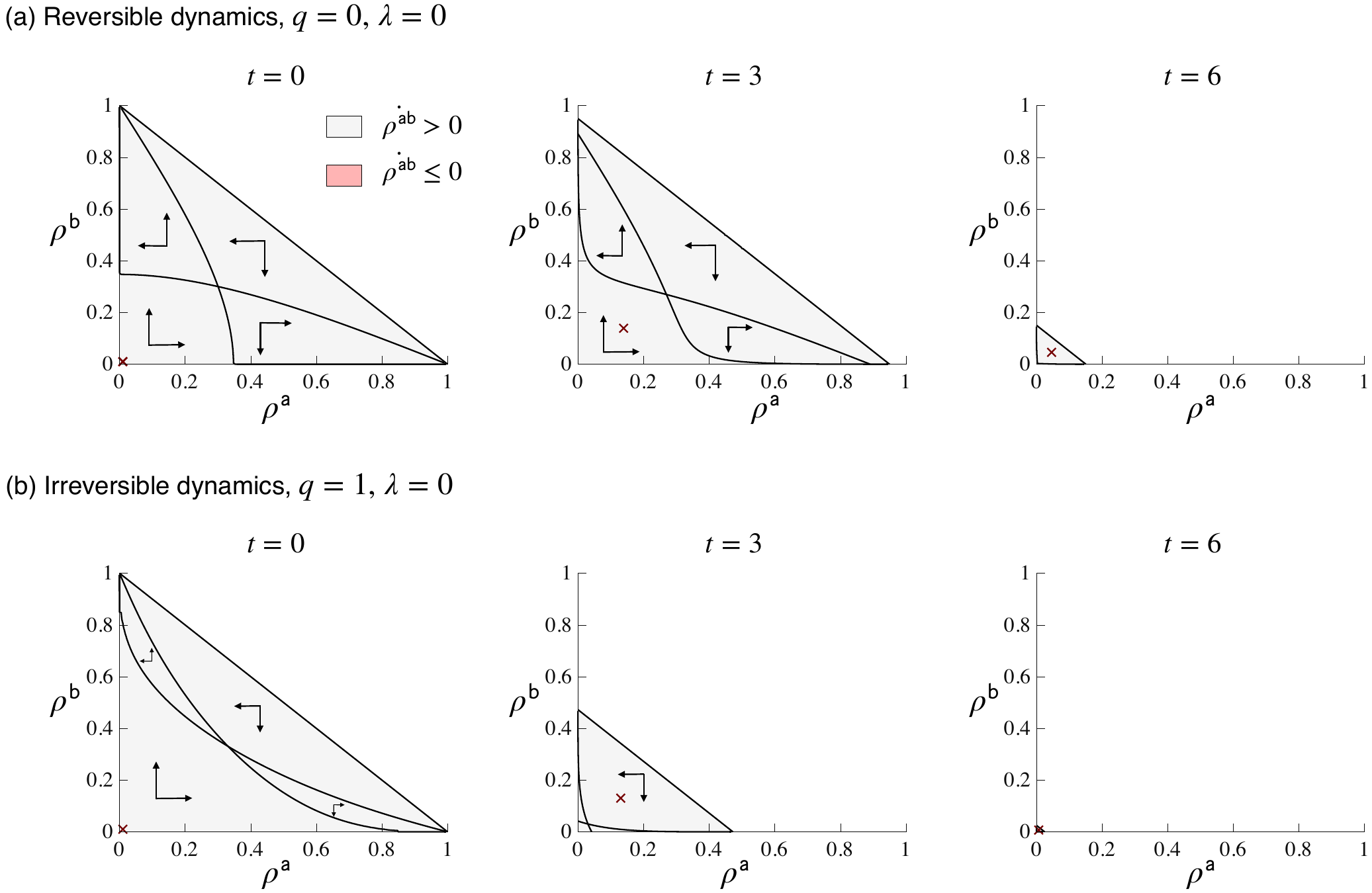}
    \caption{Phase diagrams of (a) reversible and (b) irreversible dynamics for $\lambda=0$.
    See Figs.~\ref{fig:phase_diagram_MF} and \ref{fig:irreversible_lambda-09} for the description of the phase diagrams.}
    \label{fig:irreversible_lambda0}
\end{figure}


\end{widetext}

\end{document}